\documentclass[a4paper,12pt]{article}
\usepackage{graphicx}
\usepackage{times}

\begin{document}

\title{\bf Neutron rich nuclei in density dependent relativistic
           Hartree-Fock theory with isovector mesons}

\author{B. Q. Chen\footnote{Also: Center of Nuclear Theoretical Physics,
        National Laboratory of Heavy Ion Accelarator, Lazhou 730000, P.R.China},
        Z. Y. Ma\footnote{Also: Institute of Theoretical Physics,
        Beijing 100080, P.R.China}\\
        China Institute of Atomic Energy, Beijing 102413, P.R.China\\
        \\
        F.Gr\"ummer\footnote{corresponding author: Tel: +49 (2461) 61 2802,
        Fax: +49 (2461) 61 3930,\hspace{3cm}%
        E-mail: F.Gruemmer@fz-juelich.de}
        and S. Krewald\\
        Institut f\"ur Kernphysik, Forschungszentrum J\"ulich GmbH \\
        D-52425 J\"ulich, Germany}

\maketitle

\hrule
\begin{abstract}
Density dependent relativistic Hartree-Fock theory has been extended to describe
properties of exotic nuclei.
The effects of Fock exchange terms and of $\pi$- and $\rho$- meson contributions
are discussed.
These effects are found to be more important for neutron rich nuclei than for
nuclei near the valley of stability.
\end{abstract}
\hrule

\vspace{0.5cm}
PACS: 21.10.Dr, 21.10.Gv, 21.60.Jz, 24.10.Jv\\

Keywords: relativistic Hartree-Fock, neutron rich nuclei, Ca isotopes\\

Among the various theoretical models for the bulk properties of nuclei, the
relativistic mean field theory (RMF) has been particularly
successful \cite{Ser,Rei,Gam}.
Relativistic mean field theory can be regarded as an {\sl effective field theory}
which achieves a description of nuclear properties throughout the entire periodic
table with high precision.
It is based on a relativistic Lagrangian which includes the nucleon, the omega-,
the sigma-, and the rho-meson (which reproduces the asymmetry energy of nuclear
matter) as well as non-linear self-interactions of scalar mesons.
The inclusion of the non-linear self-interactions of the sigma field introduces
density-dependent interactions in the model, which are crucial to reproduce
quantitatively the bulk properties of nuclear matter and finite nuclei.
Sharma et al. \cite{Shl}, Lalazissis et al. \cite{Lal},
Maharana et al. \cite{Mah}, Hirata el al. \cite{Hir} and
Lalazissis et al. \cite{Laf} have systematically investigated the binding
energies, deformations, rms radii, one (or two) nucleon separation energies,
quadrupole moments and isotopic shifts of nuclei far from the stability line.
A review of the relativistic mean field model has been given by Ring \cite{Ring}.

Relativistic effects have been found to be important for theories starting
from the realistic nucleon-nucleon interaction, too.
The Dirac- Brueckner-Hartree-Fock(DBHF) approach based on the meson-exchange
potential Bonn A successfully reproduces nuclear matter saturation
properties \cite{Bro}, while non-relativistic models failed.
The DBHF approach has been applied to finite nuclei by Muether et al. \cite{Mue}.
Due to the complexity of the DBHF in finite nuclei, only light spherical nuclei
have been investigated.

Brockmann and Toki have suggested a method that allows to link the binding
energies and radii of finite nuclei to the nucleon-nucleon interaction via a
local density dependence approximation to the DBHF \cite{BT}, the "relativistic
density dependent Hartree" model(RDH).
In this approach, density dependent coupling constants of the meson-nucleon
interactions are derived from DBHF calculations in nuclear matter.
Those coupling constants are used in a relativistic Hartree calculation
performed for finite nuclei.
The model therefore is parameter free in the sense that no parameters are
adjusted to data of the nuclear many-body problem.
It does, however, contain some implicit assumptions.
One assumes that it is sufficient to include density dependent modifications
of the coupling constants of the sigma- and the omega-meson only.
This assumption may be justified by the very success of the relativistic
mean field model which shows that the sigma- and the omega-meson dominate the
saturation properties of nuclei, as is obvious in closed-shell nuclei and in
the Hartree approximation.
Moreover, the Dirac Brueckner Hartree-Fock calculation provides a scalar and
a vector contribution to the self-energy of the nucleon for each density.
These are two constraints which can be used to determine two parameters of
the effective Lagrangian.
It is then natural to adjust the coupling constants of the two most relevant
mesons to those numbers.

The RDH approach with an effective sigma, omega, and rho meson gives satisfactory
descriptions of ground state properties of a few stable nuclei \cite{BT},
although the fit quality cannot match the one achieved in the RMF theories.
Investigations of this type are very important for our understanding of properties
of finite nuclei in a parameter free way.
The model so far has been applied only to a relatively small number of nuclei.
Therefore it is of great interest to extend such density dependent relativistic
model to deformed and exotic nuclei.
Recently, the RDH has been applied to deformed nuclei and the investigations
show large deviations from the experimental data for the quadrupole moments of
rare-earth nuclei \cite{Ces}.

In the RDH approach, Pauli exchange effects have been neglected.
Fritz, M\"uther and Machleidt \cite{Fri} pointed out, however,  that the nuclear
radii are reduced once the Pauli exchange effects are taken into account.
On the other hand, after incorporating both pion and rho-exchange (including
in particular the tensor part of the rho-nucleon coupling) into the
relativistic density dependent Hartree-Fock approach, larger radii are
found \cite{Ma}.
The findings of \cite{Ma} suggest that the effects of Fock terms be
investigated more thoroughly.
In the case of spin-orbit splitting, the inclusion of the Fock terms appears
to be essential.
In $^{48}$Ca, relativistic mean field theory gives a splitting of the
$d_{5/2}$ and the $d_{3/2}$ levels of 9.32 MeV, whereas the relativistic density
dependent Hartree-Fock theory gives 3.19 MeV which comes closer to the experimental
value of 4.3 MeV \cite{Ma}.

\begin{figure}[ht]
\begin{center}
\leavevmode
\includegraphics[width=10cm]{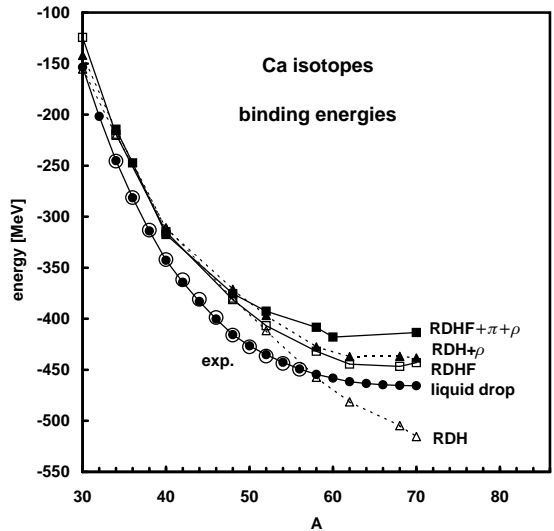}
\end{center}
\caption{\it The binding energies of the Ca isotopes are shown.
The open circles represent the experimental data \cite{Aud} and the full
line with the filled circles show a fit using a liquid drop model \cite{Moe}.
The RDH calculation including the rho meson ( dashed line with filled triangles ),
the RDHF model with sigma and omega mesons only ( full line with open squares ), and the
RDHF model including the isovector mesons ( full line with filled squares ) are compared.
An RDH calculation without isovector mesons ( dashed line with open triangles ) is given as
a reference point for the RDHF calculations.}
\label{fig1}
\end{figure}

Another place where the influence of Fock terms may be manifest is nuclei with a
large neutron excess.
The masses and the density distributions of neutron-rich nuclei are important to
predict the abundance of elements \cite{Thiel}.
Systematic experimental investigations of exotic nuclei have become possible
because of the recent advent of radioactive beam facilities \cite{Kub}.

In a study of neutron-rich nuclei, one has to include both the pion and
the rho-meson. This requires to consider the effects of short-range correlations.
Recently, Marcos et al. \cite{Marcos} have developed a nuclear matter model of
the two-body correlation function.
It provides a physical way to deal with the zero-range piece of $\pi$ and
$\rho$ interaction in the medium.
In our investigation, this effect of short-range correlations is incorporated
by removing contact interactions in $\pi$ and $\rho$ exchanges \cite{Bou}.
In addition, Marcos et al. show that the effects of short range correlations
can be simulated by modifiying the meson masses in the medium, at least
in the Hartree approximation and for vanishing momentum transfer.

In this letter, we apply the relativistic density dependent Hartree (RDH) and
Hartree-Fock (RDHF) to stable as well as  neutron rich nuclei.
We extend the calculation to include the pion and the rho-meson using the
following interaction Lagrangian:
\begin{eqnarray}
  \cal {L_{I}} & = &\bar{\psi}(g_{\sigma}\sigma
                    -g_{\omega}\gamma^{\mu}\omega_{\mu}
                    -g_{\rho}\gamma^{\mu}\rho_{\mu}\cdot\tau)\psi \\
\nonumber
               &   & +\frac{f_{\rho}}{2M}\bar{\psi}\sigma^{\mu
\nu}\partial_{\mu}
                               \rho_{\nu}\cdot\tau \psi\\ \nonumber
               &
  & -\bar{\psi}e\gamma^{\mu}A_{\mu}\frac{1}{2}(1+\tau^{3})\psi
                     -\frac{f_{\pi}}{m_{\pi}}\bar{\psi}\gamma_{5}\gamma^{\mu
}
                      \partial_{\mu}\pi\cdot\tau\psi
\end{eqnarray}
where $\tau_{i}$ indicate the isospin Pauli matrices.
The effective strengths of couplings between the nucleon and mesons are
denoted
by the coupling constants $g_{i}$ or $f_{i}$ ($i= \sigma, \omega, \rho,
\pi$),
respectively.
The formalisms of the RDH and RDHF have been presented in detail in Refs.
\cite{Fri,Boe,Ma,Bou,Sav}.

As Lenske pointed out rearrangement terms have to be included once density
dependent coupling constants are used \cite{Len}.
The inclusion of the rearrangement terms in the RDH slightly changes the
binding energy, but improves the root-mean-square radii and density distributions.
For simplicity, the rearrangement terms have not been included in this letter,
since we are interested in the contributions of the Fock term and isovector
mesons, pion and rho in neutron-rich nuclei qualitatively.

\begin{figure}[ht]
\begin{center}
\leavevmode
\includegraphics[width=10cm]{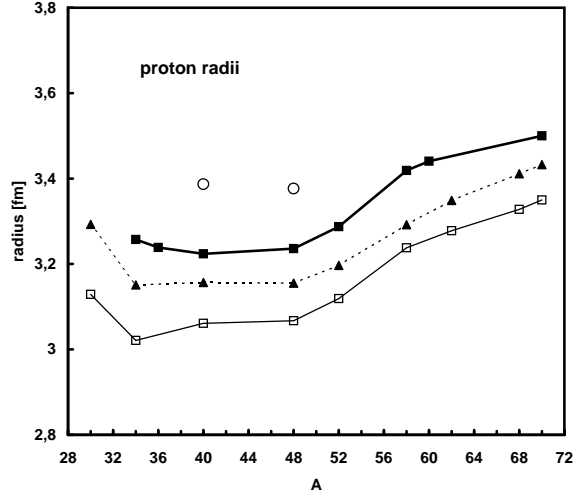}
\end{center}
\caption{\it The proton radii of the Ca isotopes are shown for some
of the models of Fig.1.
The calculations are denoted by the same symbols as in Fig.1.
In addition the open circles show the experimental proton radii of $^40$Ca and
$^70$Ca.}
\label{fig2a}
\end{figure}

For the effective interaction, the coupling constants of the sigma- and omega-
mesons are adjusted to reproduce the scalar and vector potentials U$_{s}$ and
U$_{o}$ in nuclear matter of the DBHF results using the Bonn A
potential \cite{Bro}.
The density dependence of the coupling constants is then adjusted to reproduce
the nucleon self-energies at each density in the cases of RMF or RHF with or
without isovector mesons.
The nucleon and $\omega$ meson masses are chosen to be the same as in the DBHF
calculation, where $M$ = 938.9 MeV, $m_{\sigma}$ = 550 MeV and
$m_{\omega}$ = 782.6 MeV.
For the $\pi$NN vertex pseudo-vector coupling and for $\rho$NN vector and tensor
couplings are used.
The masses and coupling constants of isovector mesons are fixed to be
m$_{\pi}$ = 138 MeV, m$_{\rho}$ = 770 MeV, $\frac{f_{\pi}^{2}} {4\pi}$ = 0.08,
$\frac{g_{\rho}^{2}}{4\pi}$ = 0.55 and $\frac{f_\rho}{g_\rho} = 3.7$ .
In the RDHF calculation with isovector mesons, the sigma meson mass has been
adjusted to $ m_\sigma = 450 $MeV ( after removal of the contact terms of the
pion and the rho meson ).
The RDHF calculation without isovector mesons has been performed with
$m_\sigma = 550 $MeV.

Our calculations are based on the one-boson exchange potential Bonn A \cite{Bro},
which is used as an input for a Dirac Brueckner Hartree-Fock calculation.
We will study the properties of exotic nuclei using three different approaches:
\begin{enumerate}
\item Relativistic density dependent Hartree theory (RDH) with the $\sigma$,
      $\omega$, and $\rho$  mesons.
\item Relativistic density dependent Hartree-Fock theory (RDHF) with $\sigma$
      and $\omega$ mesons only.
\item Relativistic density dependent Hartree-Fock theory with
      $\sigma, \omega, \pi$ and $\rho$ mesons (RDHF+$\pi$+$\rho$).
\end{enumerate}

\begin{figure}[ht]
\begin{center}
\leavevmode
\includegraphics[width=10cm]{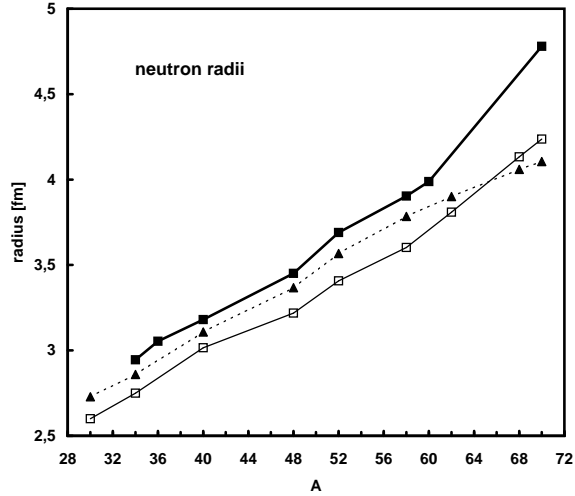}
\end{center}
\caption{\it The neutron radii of the Ca isotopes are shown for some
of the models of Fig.1.
The calculations are denoted by the same symbols as in Fig.1.}
\label{fig2b}
\end{figure}

In Fig.1, we show the binding energies of the Ca isotopes obtained in our model.
Experimentally, the binding energies of the Ca isotopes are known only till
A=56 (open circles ) \cite{Aud}.
We use the finite range liquid drop model by M\"oller, Nix and Kratz \cite{Moe}
to extrapolate the binding energies of the exotic nuclei till A=70.
The RDH approach allows to obtain a fair agreement with the experimental data,
if one incorporates the rho meson with a coupling constant
($\frac{g^2_\rho}{4\pi}=0.978)$ that has been chosen to reproduce the asymmetry
energy.

In both the relativistic density dependent Hartree- and Hartree-Fock-
approximation, the omega-nucleon and the sigma-nucleon coupling constants are
adjusted to reproduce the nucleon self energy in nuclear matter.
Therefore it is not too surprising that both methods reproduce the binding
energies of N=Z nuclei with comparable accuracy.
Indeed, Fritz et al. find in $ ^{40}$Ca a binding energy per particle of
$E/A=-8.21 MeV$ in the RDH approach, while the RDHF calculation yields
$E/A=-7.76 MeV$ which is in fair agreement with the experimental value of
-8.50 MeV.
The inclusion of the pion did not change the radius of $^{40}$Ca, but slightly
increased the binding energy \cite{Fri}.

\begin{figure}[ht]
\begin{center}
\leavevmode
\includegraphics[width=10cm]{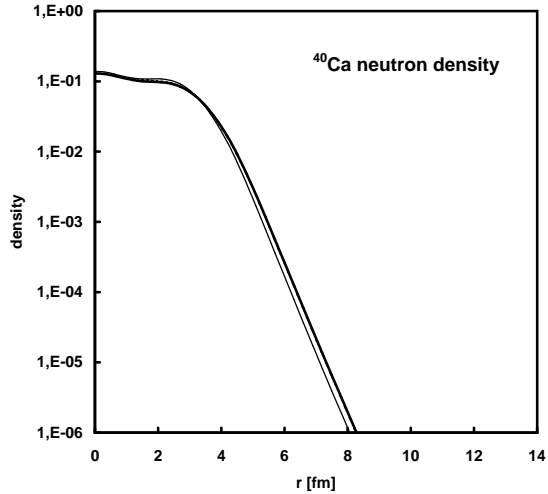}
\end{center}
\caption{\it Neutron densities  for $^{40}$Ca:
thin solid line: RDHF with $\omega $ and $\sigma $ only;
thick solid line RDHF including isovector mesons;
dashed line (hidden beneath the thick solid line) : RDH with $\rho$.}
\label{fig3a}
\end{figure}

With increasing neutron excess, however, the inclusion of the Fock terms leads
to a reduction of the binding energy.
The inclusion of the pion and the rho leads to further decreases of the binding
energies for the exotic nuclei.
The comparison with the experimental data shows that all approaches:
RDH with proper rho coupling and RDHF with and without isovector mesons could
reproduce the slope of the curve of the experimental binding energies.
At the RDHF level, the coupling constant of the rho-meson is not adjusted to the
asymmetry energy, but taken from the underlying two-body interaction.

The proton radii are displayed in Fig. 2a.
In comparison with a RDH calculation ( including the rho meson), the Fock term
reduces the proton radii for all nuclei of the isotope chain.
But the inclusion of the pion and in particular the rho-meson in the RDHF
calculation increases the proton radii.
For the neutron radii ( Fig.2b ), a reduction due to the Fock term is found for
mass numbers less than A=66, while for the extremely neutron rich nuclei, one
even finds an increase of the neutron radii.
The inclusion of the pion and the rho leads to an overall increase of the
neutron radii.

\begin{figure}[ht]
\begin{center}
\leavevmode
\includegraphics[width=10cm]{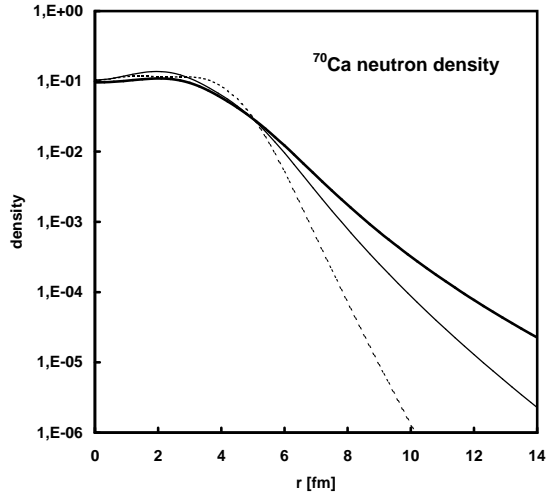}
\end{center}
\caption{\it Neutron densities  for $^{70}$Ca:
thick solid line: RDHF including isovector mesons;
thin solid line: RDHF with $\omega $ and $\sigma $ only;
dashed line: RDH with $\rho$.}
\label{fig3b}
\end{figure}

The various effects can be very clearly seen in the density distributions.
We show here as examples the neutron density distributions for $^{40}$Ca
(Fig. 3a) and $^{70}$Ca (Fig. 3b).
We see that in the case of $^{40}$Ca neither the Fock terms nor the
inclusion of isovector mesons change the neutron density distribution
considerably, so that one may argue in this case that the RDH description is
sufficiently good.
On the other hand we see strong effects of the Fock terms as well as the
isovector mesons in the case of $^{70}$Ca.
This indicates that especially in the case of very neutron rich nuclei, where
e.g. a neutron halo is expected, one has to take into account all additional
effects.
To conclude, the density dependent relativistic Hartree-Fock theory has been
employed to investigate properties of neutron rich nuclei.
For the first time the effect of Fock exchange terms and isovector mesons on
these properties has been studied.
It is found that these terms give large contributions to properties of
exotic nuclei.

This work was supported by the Deutsche Forschungsgemeinschaft (DFG) and the
China National Natural Science Foundation (CNNSF).
We thank Professor J. Speth  for constant support and helpful discussions.
B.Q. Chen and Z.Y. Ma would like to thank for the hospitality of the
Institut
f\"ur Kernphysik, Forschungszentrum J\"ulich during their visit.

\end{document}